\documentclass[showpacs,amssymb,aps,twocolumn]{revtex4}
\usepackage{amsmath}
\usepackage{graphicx}
\usepackage[latin1]{inputenc}
\begin{document}

\title{Factorization of finite temperature graphs in thermal QED}

\author{F. T. Brandt$^{a}$, Ashok Das$^{b}$, Olivier Espinosa$^{c}$,
  J. Frenkel$^{a}$ and Silvana Perez$^{d}$}
\affiliation{$^{a}$ Instituto de Física, Universidade de São
Paulo, São Paulo, BRAZIL}
\affiliation{$^{b}$ Department of Physics and Astronomy,
University of Rochester,
Rochester, New York 14627-0171, USA}
\affiliation{$^{c}$ Departamento de Física, Universidad
Técnica Federico Santa María, Casilla 110-V, Valparaíso, CHILE}
\affiliation{$^{d}$ Departamento de Física, 
Universidade Federal do Pará, 
Belém, Pará 66075-110, BRAZIL}

\bigskip

\begin{abstract}
We extend our previous analysis of gauge and Dirac fields in the
presence of a chemical potential. We consider an alternate thermal
operator which relates in a simple way the Feynman graphs in QED at finite
temperature and charge density to those at zero temperature but
non-zero chemical potential. Several interesting features of such a
factorization are discussed in the context of the thermal photon and
fermion self-energies.
\end{abstract}


\maketitle

In earlier papers \cite{silvana1,silvana2}, we gave a simple derivation of an interesting relation 
\cite{Espinosa:2003af, Espinosa:2005gq,Blaizot:2004bg} between finite temperature Feynman
graphs and the corresponding zero temperature diagrams
in the imaginary as well as in real time  formulations of
thermal field theories. We showed that in the absence of a chemical
potential, the finite temperature diagrams involving scalar, fermions
or gauge fields can be related to the zero temperature graphs through
a simple thermal operator. On the other hand, in the presence of a
chemical potential $\mu$, the thermal operator is more complicated as it
involves also time derivatives. In a theory involving fermionic fields
with a chemical potential, we have shown that a complete factorization
seems to be violated by the presence of an infrared singular
contact term \cite{silvana2}. As we have pointed out, such a behavior
may be due to the simplest possible choice of generalizing the basic
thermal operator in the presence of a chemical potential, which
involved only a single reflection operator $S(E)$ that changes the
energy as $E\rightarrow -E$.

In the present note, we propose an alternate thermal operator without
time derivatives, but which involves an additional fermionic
distribution operator $\hat N^{(T,\mu)}_F(E)$. 
We will show that such a thermal operator representation naturally
leads in QED to a simple factorization of the thermal amplitudes in the
presence of a chemical potential.

For briefness, we will discuss here the theory in the imaginary time 
formalism \cite{{kapusta:book89},{lebellac:book96},{das:book97}},
although everything also holds in the real time formalism. 
Using this formulation and the mixed space representation,
the fermion propagator at zero temperature but finite chemical 
potential can be written in the form
\begin{widetext}
\begin{equation}\label{1}
S^{(0,\mu)}(\tau,E) = \frac{1}{2E}\left[
\theta(\tau)\,A(E){\rm e}^{-(E-\mu)\tau} +
\theta(-\tau)\,B(E){\rm e}^{(E+\mu)\tau} \right],
\end{equation}
\end{widetext}
where $E=\sqrt{{\vec p\,}^2+m^2}$ and $A(E)$, $B(E)$ are Euclidean
matrices given by
\begin{equation}\label{2}
A(E) =  i\gamma^0 E - \vec\gamma\cdot\vec p + m,\;\;\;\;
B(E) = -i\gamma^0 E - \vec\gamma\cdot\vec p + m .
\end{equation}
The action of the operator $\hat N^{(T,\mu)}_F(E)$ 
on the above time dependent exponential functions
is defined as follows:
\begin{eqnarray}\label{3}
\hat N^{(T,\mu)}_F(E) {\rm e}^{s (E\pm\mu)\tau}&=&
n_F(E\pm\mu) {\rm e}^{s (E\pm\mu)\tau} 
\nonumber \\ &\equiv&
n_F^\pm(E)  {\rm e}^{s (E\pm\mu)\tau},
\end{eqnarray}
where $T$ is the temperature,
$n_F$ denotes the Fermi-Dirac distribution function and $s$ is a sign.
As shown below, this prescription enforces the anti-periodicity
condition of the thermal fermion propagator at non-zero chemical potential.

In terms of the operators $S(E)$ and $\hat N^{(T,\mu)}_F(E)$,
our alternate basic fermion operator has the simple form
\begin{equation}\label{4}
\hat{\cal O}^{(T,\mu)}_F (E) = 1-\hat N^{(T,\mu)}_F(E) \left[1-S(E)\right].
\end{equation}
Then, acting on the propagator $S^{(0,\mu)}(\tau,E)$, this fermion
operator will naturally reproduce the fermion propagator at finite
temperature and chemical potential
\begin{widetext}
\begin{equation}\label{5}
S^{(T,\mu)}(\tau,E)=\hat{\cal O}^{(T,\mu)}_F (E)S^{(0,\mu)}(\tau,E)=
\frac{1}{2E}\left\{
A(E)\left[\theta(\tau)-n_F^{-}(E)\right]{\rm e}^{-(E-\mu)\tau} +
B(E)\left[\theta(-\tau)-n_F^{+}(E)\right]{\rm e}^{ (E+\mu)\tau} \right\},
\end{equation}
\end{widetext}
which satisfies, for $-1/T<\tau< 0$, the anti-periodicity
condition \cite{KMS}
\begin{equation}
S^{(T,\mu)}(\tau,E)=-S^{(T,\mu)}(\tau+\frac{1}{T},E).
\end{equation}

The above factorization of the thermal fermion propagator is of
fundamental importance for the factorization of arbitrary graphs in QED
at finite temperature. Before we present a general proof, let us
consider first some simple amplitudes at one loop order. For example,
the integrand of
the fermion self-energy (in the Feynman gauge) may be written with the
help of Eq. (\ref{5}) in a completely factorized form as
\begin{widetext}
\begin{equation}\label{7}
\Sigma^{(T,\mu)}(E_1,E_2,\tau) = -e^2 D^{(T)}(\tau,E_1) 
\gamma_\alpha S^{(T,\mu)}(\tau,E_2) \gamma_\alpha = 
{\cal O}^{(T)}_B(E_1)\hat{\cal O}^{(T,\mu)}_F(E_2)
\Sigma^{(0,\mu)}(E_1,E_2,\tau),
\end{equation}
\end{widetext}
where ${\cal O}^{(T)}_B$ denotes the basic photon projection operator
\begin{equation}\label{8}
{\cal O}^{(T)}_B(E_1) =  1 + n_B(E_1)\left[1-S(E_1)\right]
\end{equation}
and $n_B$ is the Bose-Einstein distribution function.

If we Fourier transform Eq. (\ref{7}) in the time variable, we readily
obtain that
\begin{widetext}
\begin{eqnarray}\label{9}
&\Sigma^{(T,\mu)}(E_1,E_2,p_0) = -\displaystyle{\left.\frac{e^2}{8 E_1 E_2}\right\{ 
\left[
\frac{1+n_B(E_1) - n^{-}_F(E_2)}{E_1+E_2-\mu-ip_0} -
\frac{  n_B(E_1) + n^{-}_F(E_2)}{E_1-E_2+\mu+ip_0} 
\right] \gamma_\alpha A(E_2) \gamma_\alpha + }
\nonumber \\
&\qquad\qquad\qquad\qquad\;\;\;\;\;\;\;\;\;\;\;\;\;\;\;\;\;\;\;\;\;\,\,\,
\displaystyle{\left.\left[
\frac{1+n_B(E_1) - n^{+}_F(E_2)}{E_1+E_2+\mu+ip_0} -
\frac{  n_B(E_1) + n^{+}_F(E_2)}{E_1-E_2-\mu-ip_0} 
\right] \gamma_\alpha B(E_2) \gamma_\alpha \right\}},
\end{eqnarray}
\end{widetext}
which agrees with the result obtained by a direct evaluation. As shown
in \cite{silvana2}, apart from a contribution to the thermal mass of
the fermion, the above radiative corrections also lead to a finite
renormalization of the chemical potential which is given by
\begin{equation}\label{10}
\mu_R = \left(1-\frac{e^2}{16\pi^2}\right)\,\mu.
\end{equation}

As a second example, which exhibits a subtlety associated
with the alternate fermion operator in Eq. (\ref{4}), let us consider the
thermal photon self-energy at finite chemical potential. Using the
basic property given in Eq. (\ref{5}), 
the integrand of
this amplitude at one loop order may also be
written in a manifestly factorized form as
\begin{widetext}
\begin{eqnarray}\label{11}
\Pi^{(T,\mu)}_{\lambda\rho}(E_1,E_2,\tau) & = & e^2 {\rm Tr}\left[ 
\gamma_\lambda S^{(T,\mu)}(\tau,E_1) \gamma_\rho
S^{(T,\mu)}(-\tau,E_2)\right] 
\nonumber \\ & = & 
\hat{\cal O}_F^{(T,\mu)}(E_1) \hat{\cal O}_F^{(T,\mu)}(E_2)
\Pi^{(0,\mu)}_{\lambda\rho}(E_1,E_2,\tau) .
\end{eqnarray}
\end{widetext}
Using Eq. (\ref{1}), we see that the factors ${\rm e}^{\pm\mu\tau}$ actually
cancel in the zero temperature amplitude at finite chemical
potential. However, this would then make the action of the operator 
$\hat N_F^{(T,\mu)}(E)$ introduced in Eq. (\ref{3}) ambiguous. In 
order to define these operators unambiguously, 
one may, in the intermediate calculations, associate with each
fermion propagator of energy $E_i$ a chemical potential $\mu_i$. At
the end of the calculation, after the thermal operators have acted,
one can set $\mu_i=\mu$.

Using this procedure and Fourier transforming Eq. (\ref{11}) in the time
variable, we then readily get the result
\begin{widetext}
\begin{eqnarray}\label{12}
&\Pi^{(T,\mu)}_{\lambda\rho}(E_1,E_2,p_0) = \displaystyle{\frac{e^2}{8 E_1 E_2}}\left[ 
\frac{n_F^{-}(E_2) - n^{-}_F(E_1)}{E_2-E_1+ip_0} 
{\rm Tr} \gamma_\lambda A(E_1) \gamma_\rho A(E_2) +
\frac{n_F^{+}(E_1) - n^{+}_F(E_2)}{E_1-E_2+ip_0} 
{\rm Tr} \gamma_\lambda B(E_1) \gamma_\rho B(E_2) +
\right. \nonumber \\ 
&\qquad\qquad\qquad\qquad\;\;\;\;\;\;\;\;\;\;\;
\left. \displaystyle{\frac{1-n_F^{+}(E_2) - n^{-}_F(E_1)}{E_1+E_2-ip_0} 
{\rm Tr} \gamma_\lambda A(E_1) \gamma_\rho B(E_2) +
\frac{1-n_F^{+}(E_1) - n^{-}_F(E_2)}{E_1+E_2+ip_0} 
{\rm Tr} \gamma_\lambda B(E_1) \gamma_\rho A(E_2)} \right],
\nonumber \\ &
\end{eqnarray}
\end{widetext}
which agrees with the expression for the thermal photon self-energy
obtained by a direct evaluation. In this case, the only effect of the
chemical potential is to yield a correction to the thermal mass of the
photon \cite{lebellac:book96}. In consequence of the symmetry of thermal
QED \cite{silvana2}, the above radiative corrections do not renormalize the bare zero
chemical potential of the photon.

The complete factorization in the above examples, which occurs in
consequence of the basic relation (\ref{5}), can be immediately extended
to any one-loop graph. The difficulty in establishing such a
factorization for an arbitrary higher-loop graph arises when there are internal
vertices for which the time coordinate $\tau$ has to be integrated
over. At finite temperature, $\tau$ as well as the external times
$\tau_i$ lie in the interval $[0,1/T]$. On the other hand, in the zero
temperature graphs, the internal time needs to be integrated over the
interval $[-\infty,\infty]$. Since the basic thermal operators are
independent of the time coordinates, these may be taken out of the
time integral. Then, using the procedure outlined after Eq. (\ref{11}),
the essential step in proving the factorization
of an arbitrary graph consists in showing that the function
\begin{widetext}
\begin{equation}\label{13}
 V_\alpha = \left[\int_{-\infty}^{0} {\rm d} \tau +
                 \int_{\frac{1}{T}}^{\infty} {\rm d} \tau\right]
S^{(0,\mu_1)}(\tau_1-\tau,E_1)\gamma_\alpha
S^{(0,\mu_2)}(\tau-\tau_2,E_2) D(\tau-\tau_3,E_3)
\end{equation}
which appears in the basic electron-photon vertex, is
annihilated by the thermal operator
\begin{equation}\label{14}
{\cal O}^{(T,\mu)}(E_1,E_2,E_3) = \hat{\cal O}_F^{(T,\mu_1)}(E_1)
                                \hat{\cal O}_F^{(T,\mu_2)}(E_2)
                                    {\cal O}_B^{(T)}(E_3).
\end{equation}
To show this, we evaluate the $\tau$-integrals in Eq. (\ref{13}) and
note that the result may be written in the form
\begin{eqnarray}\label{15}
V_\alpha & = &
\left[1+{\rm e}^{-(E_1+\mu_1+E_2-\mu_2+E_3)/T}S(E_1)S(E_2)S(E_3)\right]
\nonumber \\ &  &
\displaystyle{\left[
\frac{A(E_1)\gamma_\alpha B(E_2)}{E_1-\mu_1+E_2+\mu_2+E_3}
\left(
\frac{{\rm e}^{-(E_1-\mu_1)\tau_1}}{2E_1}
\frac{{\rm e}^{-(E_2+\mu_2)\tau_2}}{2E_2}
\frac{{\rm e}^{-E_3 \tau_3}}{2E_3} . 
\right)
\right]}
\end{eqnarray}
\end{widetext}
Using the relation (\ref{3}) together with the identities
\begin{equation}\label{16}
{\rm e}^{-(E \pm\mu)/T} = \frac{n_F^{\pm}(E)}{1-n_F^{\pm}(E)};\;\;
{\rm e}^{-E_3 /T} = \frac{n_B(E_3)}{1+n_B(E_3)},
\end{equation}
it is now straightforward to show that the thermal operator given in
Eq. (\ref{14}) annihilates the quantity $V_\alpha$ in Eq. (\ref{15}). 
This establishes that for the product of propagators in the basic
electron-photon vertex, which are integrated over the common time
$\tau$, we can extend the range of integration to the interval 
$[-\infty, \infty]$.

With the help of this property, one can prove by using a 
procedure similar to the one given in 
\cite{silvana1} that in the presence of a chemical potential, an
arbitrary $N$-point thermal diagram can be factorized in the form
\begin{equation}\label{17}
\Gamma_N^{(T)} = \int \prod_{i=1}^{I}\frac{{\rm d}^3 k_i}{(2\pi)^3}
\prod_{v=1}^{V} {(2\pi)^3} \delta_v^3(k,p){\cal O}^{(T,\mu)} 
\gamma_N^{(0,\mu)},
\end{equation}
where we denote the internal and external three momenta of the graph
generically by $k$ and $p$, respectively, and
$\delta_v^3(k,p)$ enforces the three-momentum conservation at the vertex $v$.
The thermal operator for the graph is given by
\begin{equation}\label{18}
{\cal O}^{(T,\mu)} = \prod_{i=1}^{I_F} \hat{\cal O}_F^{(T,\mu_i)}(E_i)
                   \prod_{j=I_F+1}^{I}    {\cal O}_B^{(T)}(E_j),
\end{equation}
with $I_F$, $I$ being respectively the number of internal
fermion propagators and the total number of propagators.
Furthermore, $\gamma^{(0,\mu)}_N$ represents the integrand of the zero
temperature graph at finite chemical potential, which involves,
apart from a product of photon propagators, also a product
of fermion propagators $S^{(0,\mu_i)}(\tau_i,E_i)$. We note from
(\ref{1}) that, since 
$S^{(0,\mu_i)}(\tau_i,E_i)=\exp{(\tau_i \mu_i)} S^{(0,0)}(\tau_i,E_i)$,
we can moreover directly relate $\gamma^{(0,\mu)}_N$ to the integrand of the
zero temperature and chemical potential graph $\gamma^{(0,0)}_N$.
As we have pointed out, the limit $\mu_i\rightarrow \mu$ is assumed to
be taken only after the action of the thermal operator.

Thus, we have obtained a thermal operator representation for QED at
finite temperature and chemical potential, which leads to a simple
factorization of the thermal amplitudes. 
This interesting result can be also extended, by following in a
straightforward way the analysis presented in \cite{silvana2}, to a
non-Abelian gauge theory.
We would like to mention that a rather similar approach has been
recently proposed in the context of complex scalar fields at finite temperature
and charge density \cite{Inui:2006jf}. The factorization 
property is calculationally quite useful and allows
us to study in a direct and transparent way  many questions of interest at finite temperature.

\noindent{\bf Acknowledgment}

This work was supported in part by the US DOE Grant number DE-FG 02-91ER40685,
by MCT/CNPq as well as by FAPESP, Brazil and by CONICYT, Chile under grant
Fondecyt 1030363 and 7040057 (Int. Coop.).

\end{document}